\documentclass[preprint]{aastex} %one-column, single-spaced 
\shorttitle{Electron Acceleration and Time Variability} 
\shortauthors{Kusunose et al.}

\begin{document}
\title{Electron Acceleration and Time Variability \\
of High Energy Emission from Blazars}

\author{Masaaki Kusunose}
\affil{Department of Physics, School of Science, 
Kwansei Gakuin University, \\ 
Nishinomiya 662-8501, Japan}
\email{kusunose@kwansei.ac.jp}

\author{Fumio Takahara}
\affil{Department of Earth and Space Science, Graduate School of Science,
Osaka University, Toyonaka, Osaka 560-0043, Japan} 
%\email{takahara@vega.ess.sci.osaka-u.ac.jp} 

\and

\author{Hui Li}
\affil{Theoretical Astrophysics (T-6, MS B288), Los Alamos National 
Laboratory, \\ 
Los Alamos, NM 87545}
%\email{hli@lanl.gov}

\begin{abstract}
Blazars are known to emit a broad band emission from radio to 
gamma-rays with rapid time variations, particularly, in X- and 
gamma-rays.  Synchrotron radiation and inverse Compton scattering
are thought to play an important role in emission 
and the time variations are likely related to the acceleration 
of nonthermal electrons. 
As simultaneous multiwavelength observations with continuous 
time spans are recently available, some characteristics
of electron acceleration are possibly inferred from the spectral 
changes of high energy emission. 
In order to make such inferences, we solve the time-dependent kinetic 
equations of electrons and photons simultaneously using a simple model 
for electron acceleration. 
We then show how the time variations of emission are dependent 
on electron acceleration. 
We also present a simple model for a flare in X-rays and TeV 
gamma-rays by temporarily changing the acceleration timescale.
Our model will be used, in future, to analyze observed data in detail to 
obtain information on electron acceleration in blazars.
\end{abstract}


\section{INTRODUCTION}

High energy emission from blazars is usually thought to be produced by 
relativistically moving jets or blobs from the nucleus of galaxies
\citep[e.g.,][]{br78,bk79,mgc92,sbr94,it96}.
The physical properties of such jets have been probed mostly based on 
the steady state models of synchrotron radiation and inverse Compton 
scattering by a nonthermal electron population (SSC model).
However, blazars are also characterized by rapid 
and strong time variability.
Recent observations have revealed that the emission exhibits short time 
variations in X- and gamma-ray bands on timescales from weeks down to 
half an hour 
\citep[e.g.,][for review]{muk97,umu97},
as fast time variations of Mrk 421 were
observed by X-rays and TeV gamma-rays \citep{gaid96,taka96};
similar time variations of Mrk 501 were also found by multiwavelength 
observations \citep{kat99}.
These observations should provide important clues on physical processes 
in relativistic jets, in particular, on electron acceleration.  

To make theoretical inferences, we need to calculate time-dependent 
emission spectra from a time-dependent electron population.  
An example of such theoretical models was recently presented by 
\cite{mk97}.  They solved the kinetic equations of 
electrons and photons simultaneously, by injecting power-law electrons 
with an exponential cutoff.  They showed various possibilities to 
explain the time variations observed from Mrk 421, such as the changes
in the magnetic field or the maximum Lorentz 
factor of nonthermal electrons.  Although the correlation between 
X-rays and TeV gamma-rays are important to discuss the SSC model, 
their estimate of Compton scattering in the Klein-Nishina regime 
does not necessarily correctly account for the energy change in
scatterings, because of a simplified treatment of Compton scattering
in the Klein-Nishina regime
[see \cite{mk95} for the details of their calculation method].
\cite{kirketal98}, on the other hand, extended the above model, 
assuming that an acceleration region and a cooling region 
are spatially separated; 
i.e., electrons accelerated in a shock region are transferred to 
a cooling region where they emit synchrotron photons
(they did not include Compton scattering).
Their model was intended to explain the time variability of X-rays, 
by changing acceleration timescale. 

Besides the time variations of flare activities explained 
by \cite{mk97} and \cite{kirketal98}, the early stages of 
acceleration are of great importance. 
By examining the properties of the time evolutions of photon spectra during 
acceleration, we may obtain the diagnoses of acceleration mechanisms. 
The recent development of observations in X-rays (ASCA and Beppo-SAX) to 
TeV gamma-rays (e.g., Whipple and HEGRA) and future experiments 
might be used to confirm the diagnoses.

In this paper, we use a formulation similar to \cite{mk97}, 
but with the full Klein-Nishina cross section 
in the Compton scattering kernel,
so that the emission in the TeV range is calculated correctly. 
We also include a particle acceleration process 
by considering spatially separated acceleration and cooling regions 
as in \cite{kirketal98}, although we do not consider the spatial transfer
of electrons.
We particularly emphasize the detailed study of the properties of 
electron and photon spectra in the early stage of acceleration.

We describe our model in \S \ref{sec:model} and present numerical results
in \S \ref{sec:results}. Summary of our results is given in 
\S \ref{sec:summary}.

\section{MODEL}
\label{sec:model}

\subsection{Acceleration and Cooling Regions} 

We assume that observed photons are emitted from a blob moving 
relativistically towards us with Doppler factor 
${\cal D} = [ \Gamma ( 1- \beta_\Gamma \mu )]^{-1}$,
where $\Gamma$ is the Lorentz factor of the blob,
$\beta_\Gamma$ is the speed of the blob in units of 
light speed $c$, and $\mu$ is the cosine of the angle
between the line of sight and the direction of motion of the blob.
The blob is a spherical and uniform cloud with radius $R$, 
except that the blob includes an acceleration region 
which is presumably a shock front. 
It is assumed that the spatial volume of the acceleration region 
is small, and that
the acceleration region is a slab with thickness 
$R_{\rm acc}$ defined below.
The spectra of electrons and photons in the blob are calculated for 
the acceleration and cooling regions separately 
by solving equations described in \S \ref{sec:kinetic}.

We assume that the acceleration region (hereafter AR)
and the cooling region (hereafter CR) are spatially separated;
shocks in the blob are expected to be the site of electron acceleration 
and electrons cool mainly outside the shock regions. 
In the AR, electrons are mainly accelerated and 
cooling is unimportant except for the highest value of $\gamma$, 
while, in the CR, electrons with a nonthermal spectrum 
are injected from the AR;
the escape rate of electrons from the AR is equal to 
the injection rate of electrons in the CR because of the number conservation. 
We further assume that acceleration time, $t_{\rm acc}$,
and escape time, $t_{e, \mathrm{esc}}$, in the AR are energy independent 
as given in equation (\ref{eq:tacc}) below.
With these assumptions, the number spectrum of electrons 
in the AR is a power law with a power-law index $-2$, 
i.e., $N(\gamma) \propto \gamma^{-2}$, 
which is confirmed analytically \citep[e.g.,][]{kirketal98}. 
Thus the maximum energy of electrons is 
determined by the balance of cooling and acceleration. 
Since we consider $t_{\rm acc} \ll R/c$ (see \S \ref{sec:results-1}), 
the size of the AR
is much smaller than the size of the blob itself.  

We use this formulation because $N(\gamma) \propto \gamma^{-2}$
is expected from the theory of shock acceleration 
\citep[e.g.,][]{druly,be87}.
We, however, do not solve the spatial transfer of electrons 
as was done by \cite{kirketal98}.
Instead, we simply calculate escaping electrons from the AR
and put them into the CR. 
Although this may be an oversimplified model for realistic situations,
the actual geometrical situation of shocks is not well known, either.  
Strictly speaking, our formulation is valid
when ARs and CRs are more or less uniformly distributed in a cloud, 
but it is expected to be a fair approximation to the case 
where a single shock propagates in a jet as was studied by \cite{kirketal98}. 
As for the calculation of photon spectra, photons originating from one 
region penetrate into the other region but most of the photons originate 
from the CR since the size of the AR is small. 
Thus, the electron cooling in the blob is governed either 
by its own magnetic field or by synchrotron photons 
stemming from the CR. 
We treat appropriately this situation in numerical calculations.

\subsection{Kinetic Equations}
\label{sec:kinetic}

The equation describing the time-evolution of the electron number 
spectrum in the AR is given by 
\begin{equation}
\label{eq:elkinetic}
\frac{\partial N(\gamma)}{\partial t}
= - \frac{\partial}{\partial \gamma}
\left\{ \left[ \left( \frac{d\gamma}{dt} \right)_{\rm acc} 
-\left( \frac{d\gamma}{dt} \right)_{\rm loss} \right] N(\gamma) \right\} 
- \frac{N(\gamma)}{t_{e, {\rm esc}}} + Q(\gamma) \, , 
\end{equation}
where $\gamma$ is the Lorentz factor of electrons and $N(\gamma)$ is 
the number density of electrons per unit $\gamma$. 
We assume that monochromatic electrons with Lorentz factor $\gamma_0$ 
are injected in the AR, i.e., 
$Q(\gamma) = Q_0 \, \delta(\gamma-\gamma_0)$. 
Electrons are then accelerated
and lose energy by synchrotron radiation and Compton scattering;
the energy loss rate is denoted by $(d \gamma / dt)_{\rm loss}$. 
The acceleration term is approximated by 
\begin{equation}
\left( \frac{d\gamma}{dt} \right)_{\rm acc} 
= \frac{\gamma} {t_{\rm acc}} \, .
\end{equation}
In the framework of diffusive shock acceleration 
\citep[e.g.,][]{druly,be87},
$t_{\rm acc}$ can be approximated as 
\begin{equation}
t_{\rm acc} = \frac{20 \lambda(\gamma) c}{3 v_s^2} 
\sim 3.79 \times 10^{-6} \left( \frac{0.1 {\rm G}}{B} \right) \xi \, 
\gamma \quad {\rm sec},
\end{equation}
where $v_s \approx c$ is the shock speed, $B$ is the magnetic field, 
and $\lambda(\gamma) = \gamma m_e c^2 \xi / (e B)$ is the mean free path 
assumed to be proportional to the electron Larmor radius with $\xi$ 
being a parameter, $m_e$ the electron mass,
and $e$ the electron charge. Although this expression is 
valid only for test particle approximation in non-relativistic shocks, 
we rely on this since the basic dependences are not much changed in  
general cases.

For the convenience of numerical calculations, 
we assume $t_{\rm acc}$ does not depend on $\gamma$: 
\begin{equation}
t_{\rm acc} = 3.79 \times 10 \left( \frac{0.1 {\rm G}}{B} \right) 
\left( \frac{\gamma_f}{10^7} \right) \xi \quad {\rm sec}, 
\label{eq:tacc}
\end{equation}
where $\gamma_f$ is assumed to be a characteristic Lorentz factor
of relativistic electrons and used as a parameter;
we set $\gamma_f = 10^7$ throughout this paper. 
Although realistic acceleration time for the smaller values of
$\gamma$ should be correspondingly shorter, we make this choice because 
we mainly concern about the electrons with the large values of $\gamma$.
One worry about this choice is the effect on the spectrum of accelerated 
electrons.  We make sure that the resultant spectrum is that expected 
in diffusive shock acceleration by choosing 
$t_{e, \mathrm{esc}} = t_\mathrm{acc}$ in the AR;
this assumption of $t_{e, \mathrm{esc}} = t_\mathrm{acc}$ 
is the same as used by \cite{mk95}
in their proton acceleration model.

The electron spectrum in the CR is calculated by equation 
(\ref{eq:elkinetic}), with $(d \gamma / dt )_{\rm acc}$ dropped.
Also $Q(\gamma)$ is replaced by the escaping electrons from
the AR and $t_{e, {\rm esc}}$ is set to be $2 R/c$.
The assumption of $2 R/c$ in estimating $t_{e, {\rm esc}}$ 
is merely based on that electrons escaping from the blob 
take longer time than photons, and this point needs further work.

The relevant equation for the time evolution of photons is given by
\begin{equation}
\label{eq:phkinetic}
\frac{\partial n_{\rm ph}(\epsilon)}{\partial t} = \dot{n}_{\rm C}(\epsilon)
+ \dot{n}_{\rm em}(\epsilon) - \dot{n}_{\rm abs}(\epsilon) 
- \frac{n_{\rm ph}(\epsilon)}{t_{\gamma, {\rm esc}}} \, , 
\end{equation}
where $n_{\rm ph}(\epsilon)$ is the photon number density per 
unit energy $\epsilon$. Compton scattering is calculated as
\begin{equation}
\label{eq:comp}
\dot{n}_{\rm C}(\epsilon)
= - n_{\rm ph}(\epsilon) \, \int d\gamma	\, N(\gamma) \,
R_{\rm C}(\epsilon, \gamma) + \int\int d\epsilon^{\prime} \, d\gamma \, 
P(\epsilon; \epsilon^{\prime}, \gamma) \,
R_{\rm C}(\epsilon^{\prime}, \gamma) \,
n_{\rm ph}(\epsilon^{\prime}) N(\gamma) \, , 
\end{equation}
using the exact Klein-Nishina cross section.  First term of equation 
(\ref{eq:comp}) denotes the rate that photons with energy $\epsilon$ 
are scattered by electrons with the number spectrum $N(\gamma)$;
$R_{\rm C}$ is the angle-averaged scattering rate. 
Second term of equation (\ref{eq:comp}) denotes the spectrum of 
scattered photons:
$P(\epsilon; \epsilon^{\prime}, \gamma)$ is the probability that a photon 
with energy $\epsilon^\prime$ is scattered by an electron with energy 
$\gamma$ to have energy $\epsilon$. The probability $P$ is normalized 
such that $\int P(\epsilon; \epsilon^{\prime}, \gamma) \, d\epsilon = 1$. 
The details of $R_{\rm C}$ and $P$ are given in 
\cite{jones68} and \cite{bc90}.

Photon production and self-absorption by synchrotron radiation 
are included in
$\dot{n}_{\rm em}(\epsilon)$ and $\dot{n}_{\rm abs}(\epsilon)$, respectively.
The synchrotron emissivity and absorption coefficient are calculated 
based on the approximations given in \cite{rm84} for mildly
relativistic electrons
and \cite{cs86} for relativistic electrons.
External photon sources are not included. 
The rate of photon escape is estimated as 
$n_{\rm ph}(\epsilon)/t_{\gamma, {\rm esc}}$. 
We set $t_{\gamma, {\rm esc}} = R_{\rm acc}/c$ and
$R/c$ in the AR and CR, respectively,
because the scattering depth of the blob is much smaller than unity.

The comoving quantities are transformed back into the observer's frame
depending on the Doppler factor and the redshift $z$; 
$\epsilon_{\rm obs} = \epsilon \, {\cal D}/(1+z)$, 
and $dt_{\rm obs} = dt \, (1+z)/{\cal D}$

\section{RESULTS}
\label{sec:results}

We first examine the case where the cloud is initially empty and 
the injection of electrons starts at $t = 0$.
The distribution function of the injected electrons is mono-energetic:
Here $\gamma_0 = 2$ is assumed.
The strength of magnetic fields is assumed to have the same value 
both in ARs and in CRs,
which is 0.1 G except in \S \ref{sec:mag}.
Other parameters are redshift $z = 0.05$, Hubble constant 
$H_0 = 75$ km sec$^{-1}$ Mpc$^{-1}$, 
and Doppler factor ${\cal D} = 10$.
We also assume that the size of a cloud is measured 
by the timescale of variability,
which is assumed to be $R/(c {\cal D}) = 5 \times 10^4$ sec 
in the observer's frame.

\subsection{Time Evolution in Early Phase} 
\label{sec:results-1}

First we simulate the time evolution from $t = 0$ to $R/c$ to study 
the evolution in an early stage.
We assume $\xi = 5 \times 10^2$
(i.e., $t_{\rm acc} \approx 1.9 \times 10^4$ sec in the blob frame), 
and injection duration $t = 0$ -- $R/c$. 
In the CR, the escape time of electrons is assumed to be $2 R/c$.
The size of the AR is assumed to be $R_{\rm acc} = c t_{\rm acc} / 2$.
(Note that, in sections below, when we change the value of $t_{\rm acc}$, 
$R_{\rm acc}$ is also changed accordingly.) 
The injection rate of electrons in the AR is 
$0.1$ electrons cm$^{-3}$ sec$^{-1}$. 
The volume of the AR is $\sim 2.1 \times  10^{47}$ cm$^3$
and the total injection rate is 
$\sim 2.1 \times 10^{46}$ electrons sec$^{-1}$,
assuming that the AR is approximated by a disk with
radius $R$ and thickness $R_\mathrm{acc}$.
The total power of electrons amounts 
to $\sim 5 \times 10^{41}$ ergs sec$^{-1}$ by acceleration,
if the power-law spectrum with an index of 2 is realized between 
$\gamma_{\rm min}$ and $\gamma_{\rm max}$; 
the minimum and maximum Lorentz factors 
$\gamma_{\rm min}$ and $\gamma_{\rm max}$ are tentatively taken to be 
$2$ and $3\times 10^6$, respectively.

In Figure \ref{fig:el-e3-a}, the evolution of the electron number
spectrum is shown both for an AR and a CR.
It is seen that electrons injected with the Lorentz factor 2 are gradually 
accelerated and the value of $\gamma_{\rm max}$ increases with time, 
where we take the value of $\gamma_{\rm max}$ such that 
$N(\gamma) = 0$ for $\gamma > \gamma_{\rm max}$. 
The value of $\gamma_{\rm max}$ in a steady state is determined by 
the balance among $t_{\rm acc}$, $t_{e, {\rm esc}}$,
and cooling time $t_{\rm cool}$ in the AR.
Because we assume $t_{\rm acc} = t_{e, {\rm esc}}$ 
in the AR, $\gamma_{\rm max}$ is simply 
determined by $t_{\rm acc}$ and $t_{\rm cool}$. 
The value of $\gamma_{\rm max}$ in Figure \ref{fig:el-e3-a}
is $\sim 4 \times 10^6$.
The spectrum reaches almost a steady state within $R/c$, 
which is a power law, $N(\gamma) \propto \gamma^{-2}$;
note that $t_{\rm acc} \sim 2 \times 10^4$ sec
and $R/c = 5 \times 10^5$ sec 
in the comoving frame of the blob for the present model.

In the CR, the effect of electron escape is negligible in 
the time interval shown in Figure \ref{fig:el-e3-a}, because 
the simulation is terminated 
at $t = R/c$ while $t_{e, {\rm esc}} = 2 R/c$.
Because of radiative cooling, 
a break $\gamma_{\rm br}$ appears at around $3 \times 10^5$.
This break moves to lower energy when the evolution is continued 
until a steady state is attained; 
$\gamma_{\rm br} \sim 10^4$ at $t = 10 R/c$. 
There is also a slight deceleration of electrons by cooling, 
which is shown by curves below $\gamma = 2$.

The spectral energy distribution (SED) of emission from the CR
is shown in Figure \ref{fig:ph-e3-a};
the flux and the photon energy are plotted in the observer's frame.
Curves in the figure show the time evolution, 
with equally spaced time interval for $t = 0 - R/c$ by solid curves. 
They evolve from lower to upper curves.
In this stage, the synchrotron radiation dominates, because 
Compton scattering needs a timescale $\sim R/c$ to be effective. 
SED at $t = 2 R/c$ (dotted curve) and $10 R/c$ (dashed curve) are also 
shown in the figure; here electrons are continuously injected
until $t = 10 R/c$.
As shown by those curves, when the evolution is continued after $R/c$, 
the Compton component continues to increase before reaching
a steady state.
The peak energy of synchrotron emission initially increases 
but begins to decrease after about $0.8R/c$ 
because electrons with $\gamma < \gamma_{\rm br}$ 
continue to accumulate and the value of $\gamma_{\rm br}$
decreases while those with $\gamma > \gamma_{\rm br}$ 
are saturated because of radiative cooling.
After $t = 2R/c$ the effects of electron escape begin to further
modify the synchrotron spectrum; the intensity 
at the high energy part decreases 
while that at low energy still continues to increase slightly. 

For $t = 0 \sim R/c$, light curves in the X-ray range are
shown in Figure \ref{fig:lightcv-e3-a}.
Hard X-rays become dominant after $t \sim 15 t_\mathrm{acc}$,
where $t_\mathrm{acc}/{\cal D} \sim 2 \times 10^3$ sec 
in the observer's frame.

The time evolution of the energy densities of electrons and photons 
in the CR are shown in Figure \ref{fig:ene-e3}:
The energy densities in the AR are comparable with those 
in the CR for the parameters we used.
In the CR, $t_{e, {\rm esc}} = 2 R/c$ and 
$t_{\gamma, {\rm esc}} = R/c$ are assumed, 
so that the energy density of electrons is larger than that of photons. 
As was mentioned above, first the synchrotron photon energy-density 
rapidly increases and later the Compton photon 
energy-density (indicated by SSC in the figure) increases.
It should be noted that the ratio of the energy densities 
of the Compton component to the synchrotron component
is about 0.7 in the final stage, 
while the ratio of energy densities of synchrotron photons to 
magnetic fields is about 9. 
This is because the energy range of the target photons of Compton scattering 
is only a part of the synchrotron spectrum due to the Klein-Nishina limit. 
This result implies that we should be cautious about the estimate of 
the magnetic field strength from observations; 
if we simply estimate the magnetic field by multiplying 
the energy density of synchrotron photons 
by the ratio of synchrotron luminosity to Compton luminosity, 
it results in a large overestimation of magnetic field. 

The energy injected through the electron acceleration is finally
carried away by electrons and photons from the blob.
The ratio of the amounts of the energies
carried by electrons and photons is about $1.8 : 1$
in a steady state (i.e., $t \sim 10 R/c$).
That is, electrons carry more jet power than radiation
in this specific model.

The trajectories in the energy-flux {\it vs.} photon-index plane are 
shown for $t = 0$ -- $10 R/c$ in Figure \ref{fig:alpha-e3} 
for various energy bands.
Because the value of $\gamma_{\rm max}$ decreases due to
radiative cooling, 
the flux of X-rays decreases when $t$ $>$ a few $R/c$.
The flux of gamma-rays, on the other hand, continues to increase
because of Compton scattering (see Figure \ref{fig:ph-e3-a}).

\subsection{Dependence on Acceleration Timescale} 

By changing the value of $\xi$,
we compare the electron spectrum for different values of the 
acceleration time,
where we keep $t_{\rm acc} = t_{e, {\rm esc}}$ in the AR.
In Figure \ref{fig:el-e3-8-9},
steady-state distributions of electrons
for different values of $\xi$ are compared. 
When the acceleration time scale is longer, 
the value of $\gamma_{\rm max}$ is reduced because of 
radiative cooling in the AR. 
Consequently, the emission spectrum becomes softer 
(Figure \ref{fig:ph-e3-8-9}).
It should be noted that because a smaller value of $\xi$
leads to a smaller value of $R_\mathrm{acc}$ in out model,
the luminosity from a blob becomes smaller when $\xi$ is smaller.
The extreme limit of $\xi=1$ corresponding to the B\"ohm limit 
results in the most efficient Compton luminosity and the highest 
gamma-ray energy. 
In this limit, $\gamma_{\rm max}$ is about $2 \times 10^9$ 
and the inverse Compton SED shows a steep cut off at $\sim 10^4$ TeV
for ${\cal D} = 10$ if electron-positron pair production is neglected.
Note that $t_{\rm acc}$ in reality depends on $\gamma$,
while we assume $t_{\rm acc}$ does not depend on $\gamma$
and the above values were calculated assuming $\gamma_f = 10^7$ in
equation (\ref{eq:tacc}).

The shape of SED has a significant curvature in the TeV region
in our calculations.
This curvature is in contrast to the observations of 
TeV gamma-rays from Mrk 421,
which are fitted by a power law \citep{kretal99}.
Mrk 501, on the other hand, show a curvature in TeV emission
\citep{cat97}, and there are models which explain the curvature
by intergalactic absorption \citep[e.g.,][]{ko99,kretal99}.
We, however, do not address these issues in this paper,
since we are mainly interested in the temporal behavior of
electrons and photons due to electron acceleration in the source.

\subsection{Dependence on the Injection Rate}

The spectral energy distributions of electrons and photons
depend on the value of the injection rate 
$Q(\gamma)$ in the AR as well.
If the value of $Q(\gamma)$ is larger
with the fixed values of $\gamma_0$, $t_{e, {\rm esc}}$, and $t_{\rm acc}$,
the accumulation of electrons in the CR increases, 
resulting in the dominance of the Compton component. 
An example of SED is shown \ref{fig:ph-e4-e3},
where the electron injection rate in the acceleration 
smaller by a factor 10 than in the model shown in Figure \ref{fig:ph-e3-a},
i.e., electrons are injected at the rate of $0.01$ electrons
cm$^{-3}$ sec$^{-1}$. 
The peak of the synchrotron component decreases by a factor 10
and that of the Compton component decreases by a factor 100.

\subsection{Dependence on Magnetic Field}
\label{sec:mag}

When the size of a cloud and the number density of electrons are
fixed, the value of $\gamma_{\rm max}$ is larger for smaller values of $B$,
because the synchrotron cooling rate is proportional to $B^{2}$.
However, this is not the case in our model,
because not only the cooling rate but also $t_{\rm acc}$ depends on B.
When $B$ is smaller, $t_{\rm acc}$ is larger,
which results in the larger size of the AR.
Because we fix the particle injection rate per unit volume in the AR, 
the total number of electrons injected
into the CR per unit time is larger by
the electron number conservation.
As a result, the Compton cooling in the CR
becomes stronger and the value of $\gamma_{\rm max}$ becomes smaller.
However, the increase or decrease of $\gamma_{\rm max}$
actually depends on the combination of synchrotron cooling
and Compton cooling.
Such dependence on $B$ in the CR is shown in Figure \ref{fig:light-b05-g2};
SEDs at $t = 10 R/c$ are compared for $B = 0.05$, $0.1$, and
$0.5$ G with the same values of other parameters as
in Figure \ref{fig:ph-e3-a}. 
In the CR, the values of $\gamma_{\rm max}$ are $8 \times 10^6$,
$4 \times 10^6$, and $8 \times 10^5$ for 
$B = 0.05$, $0.1$, and $0.5$ G, respectively.

\subsection{Termination of Acceleration} 

It is conceivable that acceleration is terminated 
by the end of electron injection in the AR
due to the change of shock structure, etc., so that 
plasmas cease to emit hard photons.  
To exemplify such a situation, 
we continue the injection and acceleration up to $t = 4 R/c$ with the 
parameters used in Figure \ref{fig:el-e3-a} and terminate the injection 
and the acceleration abruptly at $t = 4 R/c$, 
while the simulation is continued until $t = 7 R/c$. 
A break of the power-law spectrum of electrons 
in the AR appears after acceleration is terminated,
and the break moves to lower energy with time. 
The response of the emission spectrum to the termination of 
acceleration is almost simultaneous
in different energy bands as shown by light curves 
in Figure \ref{fig:light-e3-c}. 
It is observed that the decay at 0.5 -- 2 keV band lags 
that at 2 -- 40 keV,
which is characteristic to the models that assume
the injection of power-law electrons and a sudden termination
of injection.
The decay in the keV range and 1 -- 10 TeV bands is exponential,
because the supply of the electrons producing those photons 
is turned off.
On the other hand, electrons producing GeV photons are still supplied for
a while by the cooling of the highest energy electrons 
which produced 1 -- 10 TeV photons.

\subsection{Flare}

Up to now, we have assumed that at the initial stage 
the cloud is empty and there are no high energy electrons or photons. 
This is certainly an over simplification. 
Many flare events have been observed in X- and gamma-ray ranges 
by ASCA, Whipple, etc. They are overlaid on a steady emission component. 
As an example of applications of our code, a flare is simulated, 
i.e., we simply change the value of 
$t_{\rm acc}$ for a period of time. 
More specifically, at $t = 0$ the distributions of electrons 
and photons are in the steady state which is obtained
for the parameters used in \S \ref{sec:results-1};
see the dashed curve in Figure \ref{fig:ph-e3-a} 
for the steady photon energy distribution.
The steady state is still continued for $R/c$.
We then replace $t_{\rm acc}$ by
$t_{\rm acc}/1.2$ for $t = R/c - 2 R/c$
(about 14 hours in the observer's frame); 
after $t = 2 R/c$, the original value of $t_{\rm acc}$ is used. 
The electron escape time in the AR is 
also changed keeping $t_{e, {\rm esc}} = t_{\rm acc}$. 
In Figure \ref{fig:light-flare}, light curves are shown for
such a flare.
The response of the light curves to the change of 
$t_{\rm acc}$ (on/off of a flare) is slightly delayed, 
because of photon production and Compton scattering time. 
It is also noticed that the change of the light curve 
at $1 - 10$ GeV band delays behind X-rays and TeV gamma-rays. 
This is a result of an interplay of the time evolution of 
electron and synchrotron photon spectra. 
It is shown that the light curves of 2 -- 10 and 10 -- 40 keV
proceed that of 0.5 -- 2 keV.
This behavior is different from that shown 
in Figure \ref{fig:lightcv-e3-a},
where the initial condition was an empty blob.

The trajectories in the energy flux and photon index are shown 
in Figure \ref{fig:alpha-flare} for $t = 0 $ -- $10 R/c$. 
This behavior is qualitatively similar to observed one for Mrk 421 by 
ASCA \citep{taka96}.
Though the amplitudes of the change in the photon index of 2 -- 10 keV 
and its energy flux are different from those of the observation,
these values are dependent on parameters such as $t_{\rm acc}$ 
and the duration of the flare, etc.

\section{SUMMARY}
\label{sec:summary}

Simulations of the time evolution of electron and photon energy 
distributions were presented as a model of time variations observed 
by X- and gamma-rays from blazars. 
By assuming that acceleration and cooling regions in a blob are 
spatially separated,
we calculated the energy spectra of electrons in each regions. 
Electrons in the acceleration region are accelerated with a characteristic 
timescale $t_{\rm acc}$ and escape on a timescale $t_{e, {\rm esc}}$; 
here we assumed $t_{\rm acc} = t_{e, {\rm esc}}$, so that the electron 
spectrum in a steady state obeys a power law,
$N(\gamma) \propto \gamma^{-2}$, as realized in the standard model 
of shock acceleration \citep[e.g.,][]{druly,be87}.
Electrons escaping from the acceleration region are injected into 
the cooling region
where they lose energy by radiation and finally escape from the blob 
on a timescale assumed to be $2R/c$.
With these assumptions, we performed the simulations
of the time evolutions of electrons and photons for various 
values of parameters.
Although we did not include a specific acceleration mechanism, 
we took into account the salient features of diffusive shock
acceleration, so that we could study the properties of time variation
accompanying shock acceleration.

We first presented the results of
the time evolution of the spectral energy distribution of radiation
associated with the evolution of the electron number spectrum.
In the early stage of the evolution, i.e., 
$t = 0$ -- $R/c$, the synchrotron component dominates the spectrum.
The energy flux of soft X-rays starts to rise earlier than that
of hard X-rays.
Later ($t > R/c$), the Compton luminosity gradually increases.
At the same time, the peak energy of the synchrotron component
decreases because of radiative cooling.
It was found that in a steady state, 
escaping electrons carry more energy than radiation:
This result, of course, depends on the values of the parameters used.
We also showed the dependence of time evolution on 
the acceleration timescale,
the electron injection rate, and the strength of magnetic fields.
The value of $\gamma_{\rm max}$ and the ratio of the synchrotron
luminosity to the Compton luminosity depend on such parameters.

We next simulated a flare by simply changing the value of 
$t_{\rm acc}$ for a certain time span.
With a shorter acceleration timescale,
more energetic electrons are produced and consequently 
more hard photons are produced.  
The relation between the energy flux and the photon index during a flare
was obtained, which is similar to the one 
observed from Mrk 421 \citep{taka96}.

Our formulation provides a method to treat high energy flares 
including particle acceleration processes, which is beyond 
usual analyses where nonthermal electron spectra are arbitrarily assumed 
and only cooling processes are included.
Although we have not applied our model to any specific case of 
flares, it is straightforward to do this using our code. 
The examples presented here seem to cover a wide range of 
observed flares. 
These applications are deferred to future work.  
On the theoretical side, as proposed by \cite{kirketal98},
electrons accelerated at a shock are transferred outside of
the shock and cool radiatively. 
To include such spatial transfer of electrons,
we, in future, need to solve for the structure 
around acceleration regions.

Recently \cite{cg99} showed observational consequences
associated with time variations with timescales shorter than $R/c$.
When such short timescale variations occur,
observed emission is a superposition from various parts of a cloud.
Then the time profile of each time variation is not necessarily
observed clearly.
The model presented in this paper contains the acceleration timescale
shorter than $R/c$.
Thus our model may not directly reflect observed spectra.
However, to understand the relation between electron acceleration
and time variation of emission, such a study should be useful.

\acknowledgements

M.K. and F.T. have been partially supported by Scientific Research Grants 
(M.K.: Nos. 09223219 and 10117215; F.T.: Nos. 09640323, 10117210, and 
11640236) from the Ministry of Education, Science, Sports and Culture 
of Japan.

%\clearpage

\clearpage
  
% Figure 1
%\epsfig{file=fig1.ps,height=16cm,width=15cm} 
\scalebox{0.9}[0.9]{\includegraphics{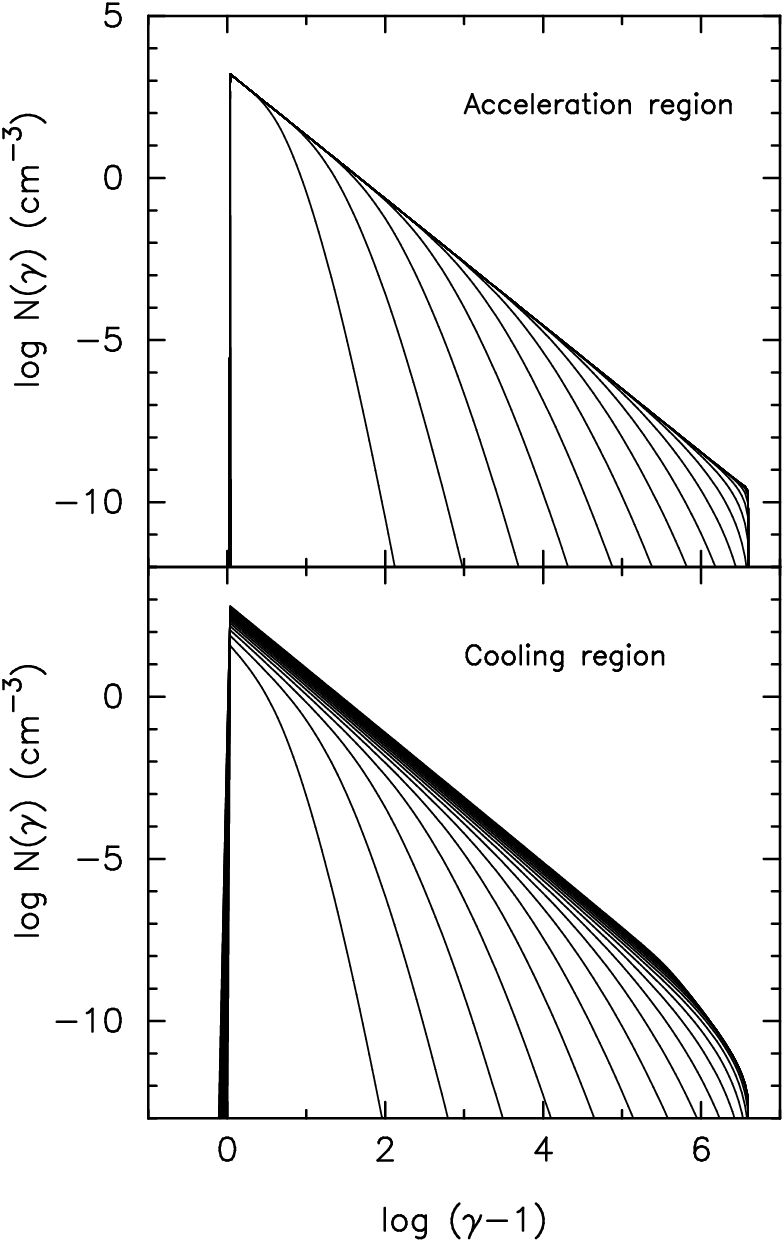}}

\figcaption{
Time evolution of electron number spectra 
in the acceleration (upper panel) and cooling (lower panel)
regions for $t = 0$ -- $R/c$ 
with the equally spaced time span of $0.05 R/c$.
The spectra evolve from lower to upper curves in each panel.
\label{fig:el-e3-a}
}

% Figure 2
% \clearpage
%\epsfig{file=fig2.ps,height=15cm,width=15cm} 
\scalebox{0.9}[0.9]{\includegraphics{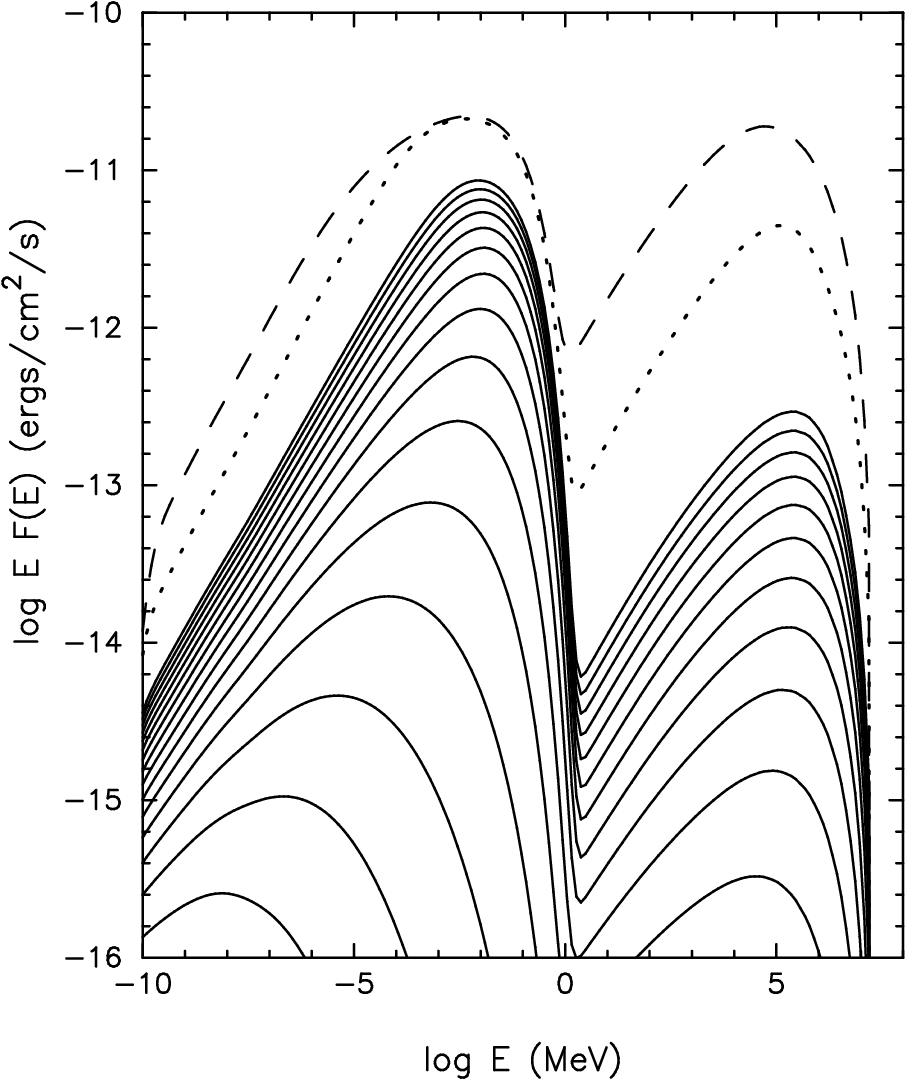}}

\figcaption{
Time evolution of the spectral energy distribution (SED) 
of photons emitted by electrons in the cooling region shown 
in Figure \ref{fig:el-e3-a};
SED is shown in the observer's frame.
The solid curves are for $t = 0$ -- $R/c$ (lower to upper curves)
with the equally spaced time span of $0.05 R/c$.
SEDs when the simulation is continued after $R/c$ 
with continuous injection and acceleration are also shown; 
the dotted curve shows SED at $t = 2R/c$ 
and the dashed curve is SED at $t = 10R/c$, 
at which the radiation is already in a steady state. 
\label{fig:ph-e3-a}
}
% wip .. Work-osaka3/Fig-e3-a/photon.wip %

% Figure 3
% \clearpage
%\epsfig{file=fig3.ps,height=15cm,width=15cm} 
\scalebox{0.9}[0.9]{\includegraphics{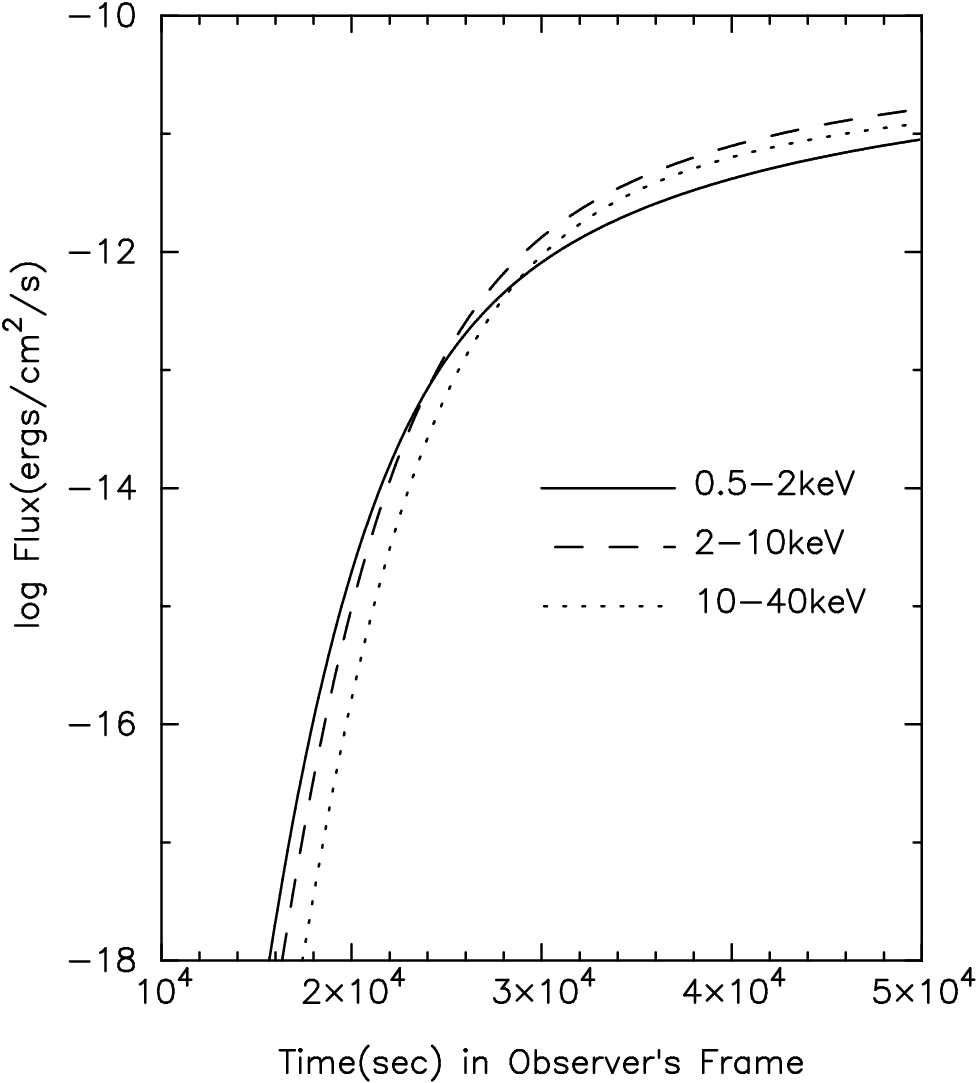}}

\figcaption{Light curves for 0.5 -- 2, 2 -- 10, and 10 -- 40 keV
bands.  The energy flux of soft X-rays is larger than
that of hard X-rays for $t \lesssim 15 t_{\rm acc}$.
\label{fig:lightcv-e3-a}
}

% Figure 4
% \clearpage
%\epsfig{file=fig4.ps,height=15cm,width=15cm} 
\scalebox{0.8}[0.8]{\includegraphics{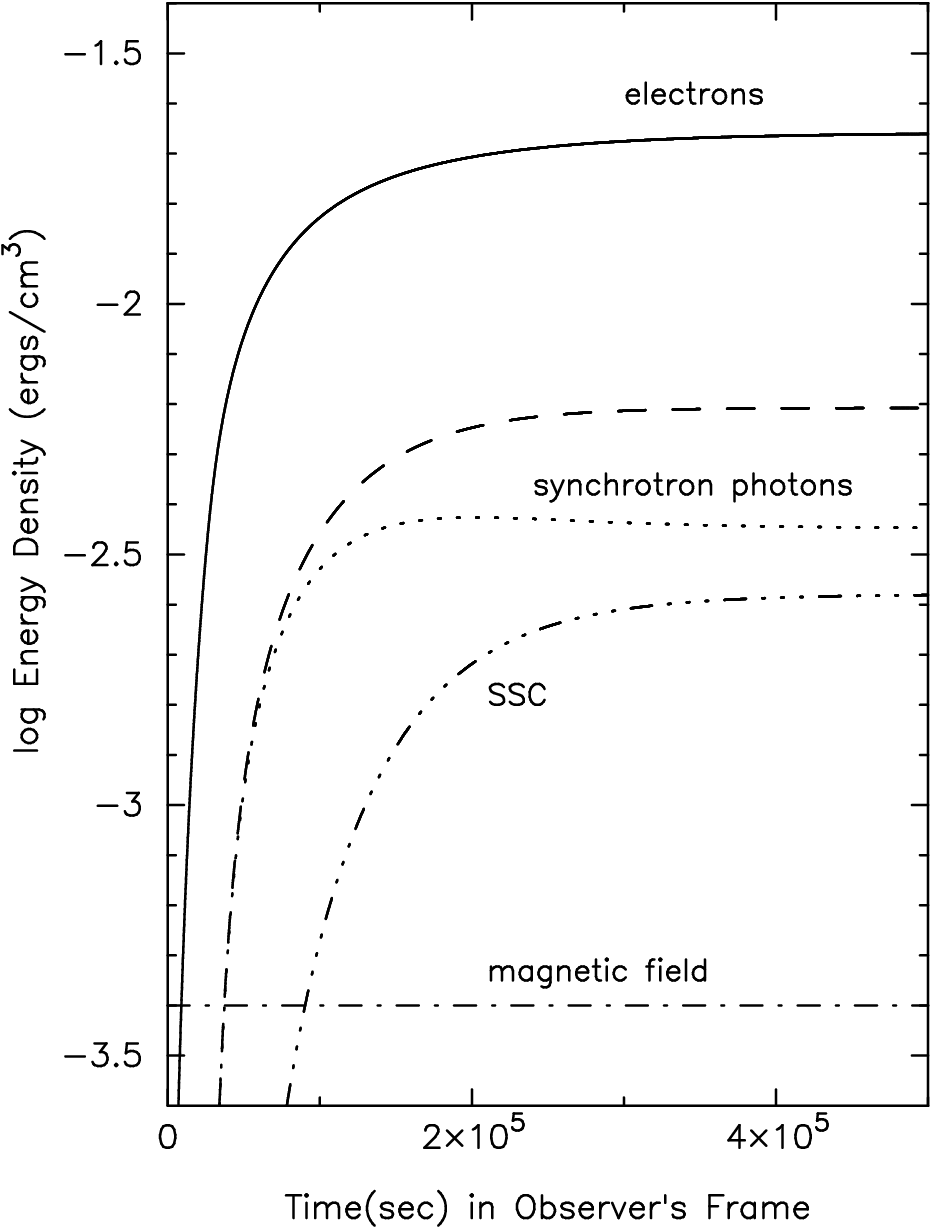}}

\figcaption{
Time evolution of the energy densities of electrons and photons
in the cooling region.
Here the strength of magnetic field is fixed, $B = 0.1$ G.
Curves are plotted for $t = 0$ -- $10 R/c$,
where $R/(c {\cal D}) = 5 \times 10^4$ sec 
and $t_{\rm acc}/{\cal D} \approx 2 \times 10^3$ sec in the observer's frame.
Solid curve: electrons, long-dashed: photons (synchrotron plus SSC),
dotted: synchrotron photons, dash-dot-dot-dotted: SSC photons,
and dash-dotted: magnetic field.
\label{fig:ene-e3}
}

% Figure 5
% \clearpage
%\epsfig{file=fig5.ps,height=15cm,width=15cm} 
\scalebox{0.9}[0.9]{\includegraphics{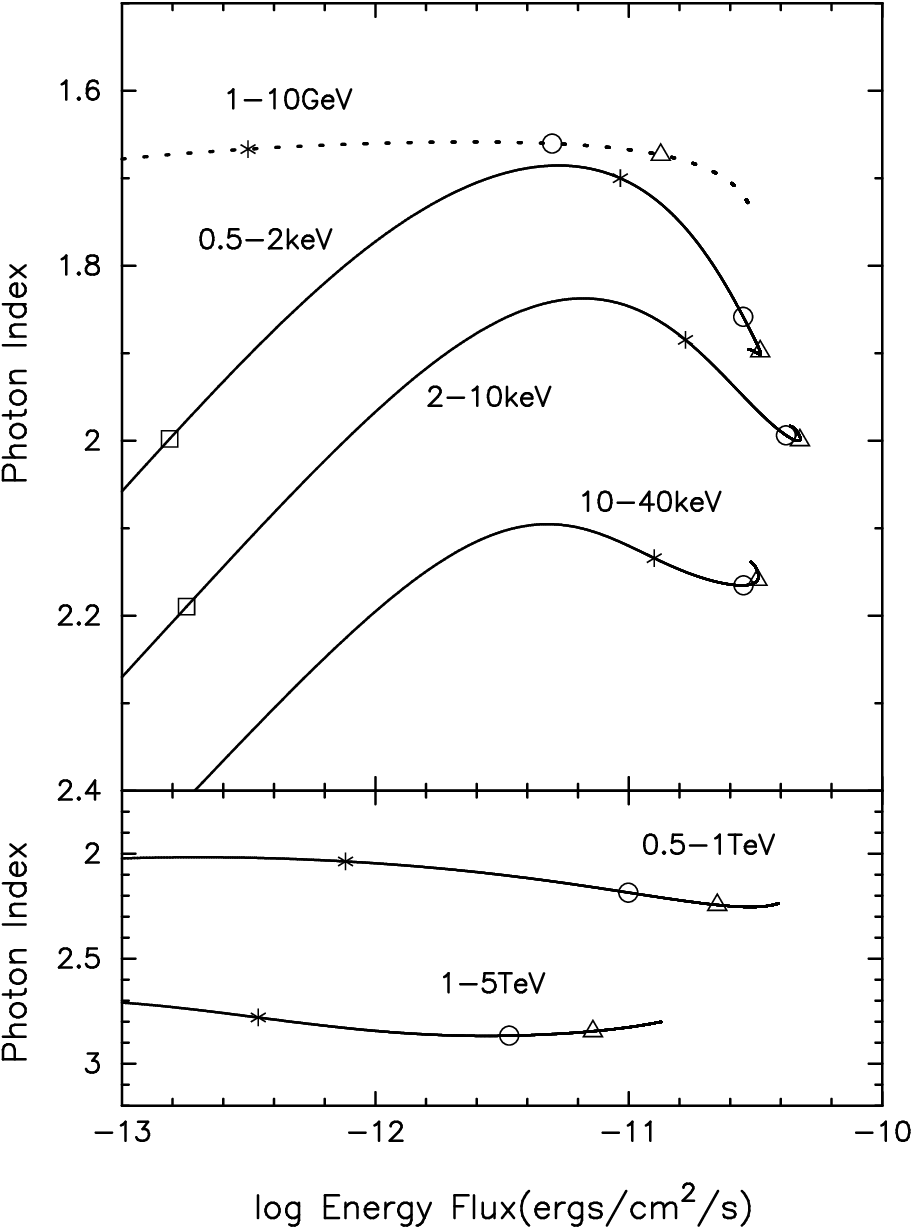}}

\figcaption{
Trajectory in the energy-flux and photon-index plane for
various energy bands.
The evolution is calculated for $t = 0$ -- $10 R/c$.
Symbols on the curves indicate the time from $t = 0$; 
$t = 0.5 R/c$ (squares), $R/c$ (asterisks), 
$2 R/c$ (circles), and $3 R/c$ (triangles).
\label{fig:alpha-e3}
}

% Figure 6
% \clearpage
%\epsfig{file=fig6.ps,height=18cm,width=15cm} 
\scalebox{0.9}[0.9]{\includegraphics{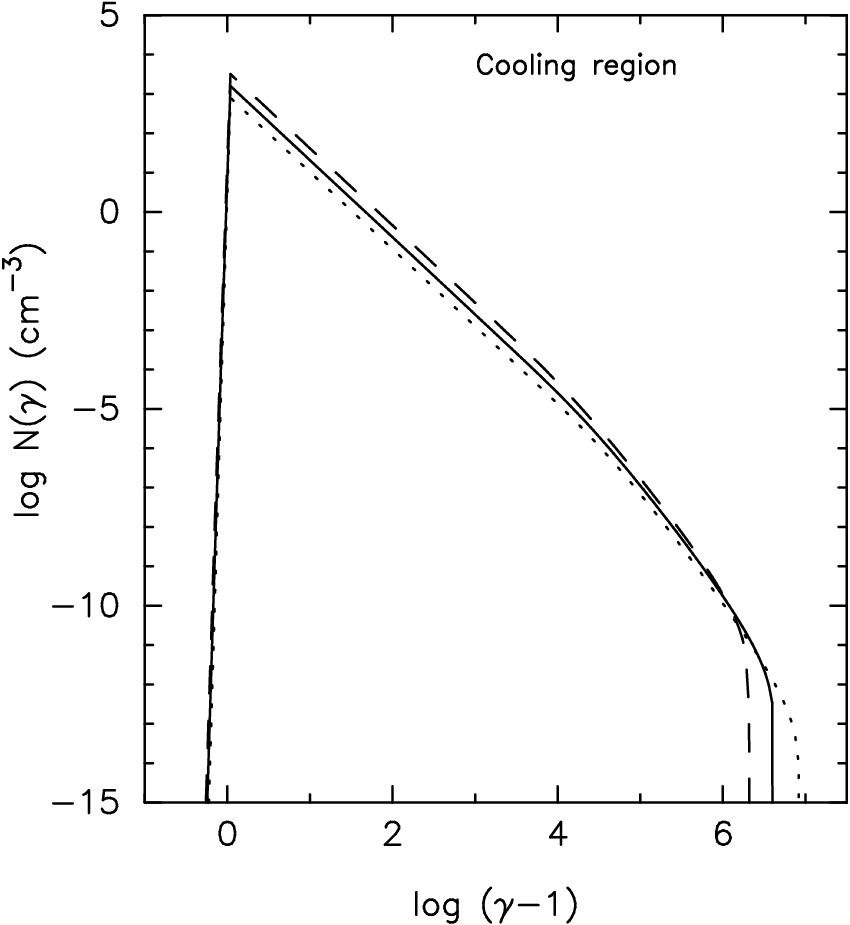}}

\figcaption{
Electron distribution in the cooling region at $t = 10 R/c$ for 
various values of the acceleration timescale.
The solid curve is for $\xi = 5 \times 10^2$,
the dashed curve for $\xi = 10^3$,
and the dotted curve for $\xi = 2.5 \times 10^2$. 
\label{fig:el-e3-8-9}
}

% Figure 7
% \clearpage
%\epsfig{file=fig7.ps,height=18cm,width=15cm} 
\scalebox{0.9}[0.9]{\includegraphics{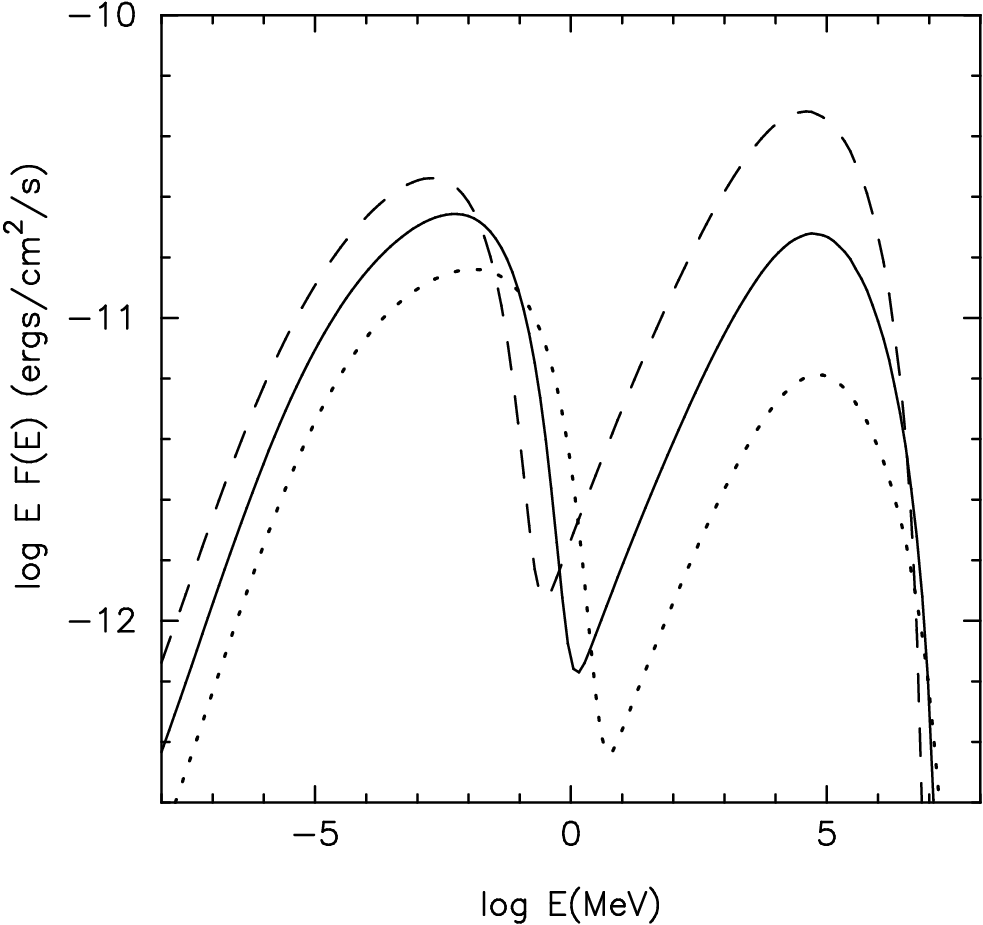}}

\figcaption{
SDEs at $t = 10 R/c$, corresponding 
to Figure \ref{fig:el-e3-8-9}, for various values of 
the acceleration timescale.  The solid curve is for $\xi = 5 \times 10^2$,
the dashed curve for $\xi = 10^3$,
and the dotted curve for $\xi = 2.5 \times 10^2$. 
\label{fig:ph-e3-8-9}
}

% Figure 8
% \clearpage
%\epsfig{file=fig8.ps,height=16cm,width=15cm} 
\scalebox{0.9}[0.9]{\includegraphics{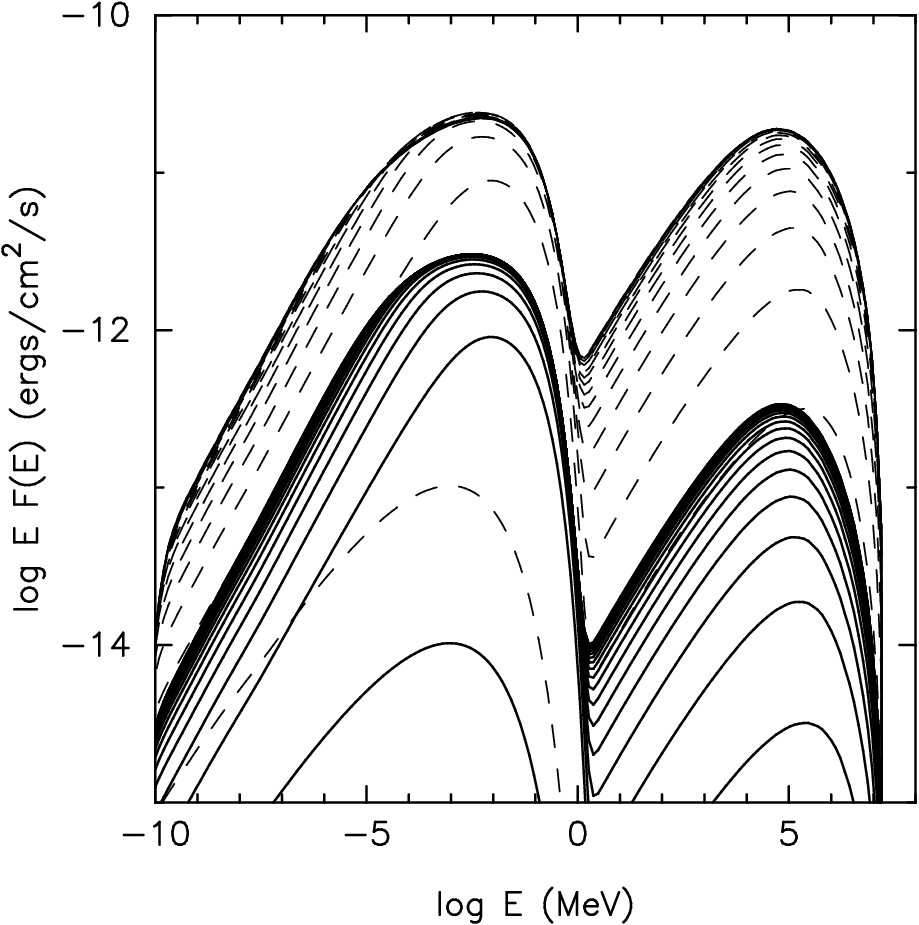}}

\figcaption{
Evolution of SEDs for different values of $Q(\gamma)$.
The solid curves are the evolution of SED with $Q(\gamma)$ smaller
by a factor 10 than that of Figure \ref{fig:ph-e3-a} 
shown here by the dashed curves. 
The curves are plotted for $t = 0$ -- $10 R/c$ with
the time interval $0.5 R/c$ and evolve from lower to upper.
\label{fig:ph-e4-e3}
}

% Figure 9
% \clearpage
%\epsfig{file=fig9.ps,height=16cm,width=15cm} 
\scalebox{0.9}[0.9]{\includegraphics{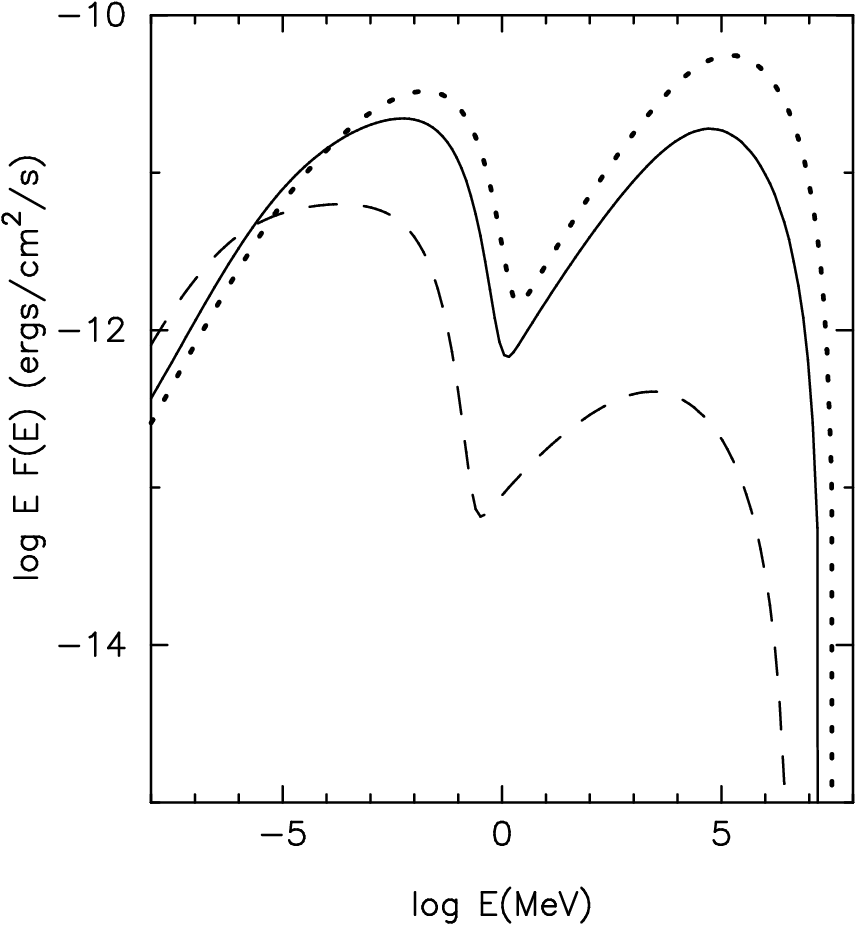}}

\figcaption{
SED at $t = 10 R/c$ for different values of $B$.
The solid curve is SED for $B = 0.1$ G (the same curve as shown
in Figure \ref{fig:ph-e3-a}),
the dotted curve is SED for $B = 0.05$ G,
and the dashed curve is for $B = 0.5$ G.
\label{fig:light-b05-g2}
}

% Figure 10
% \clearpage
%\epsfig{file=fig10.ps,height=18cm,width=14cm} 
\scalebox{0.8}[0.8]{\includegraphics{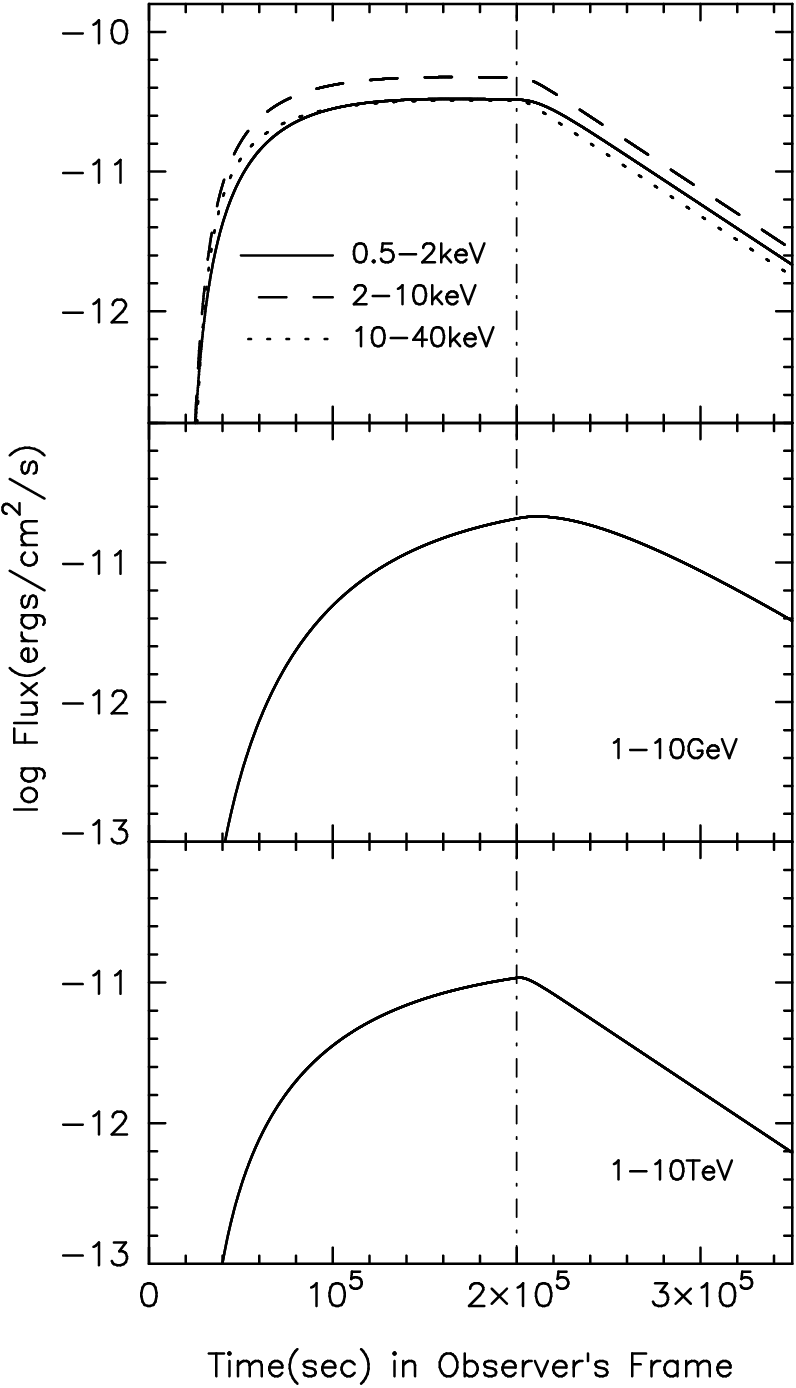}}

\figcaption{
The response of light curves to the termination of acceleration. 
Acceleration and injection are terminated at $4 R/c$ 
or $2 \times 10^5$ sec in the observer's frame,
shown by the vertical dash-dotted line.
Parameters are the same as in Figure \ref{fig:el-e3-a}. 
\label{fig:light-e3-c}
}

% Figure 11
% \clearpage
%\epsfig{file=fig11.ps,height=18cm,width=15cm} 
\scalebox{0.8}[0.8]{\includegraphics{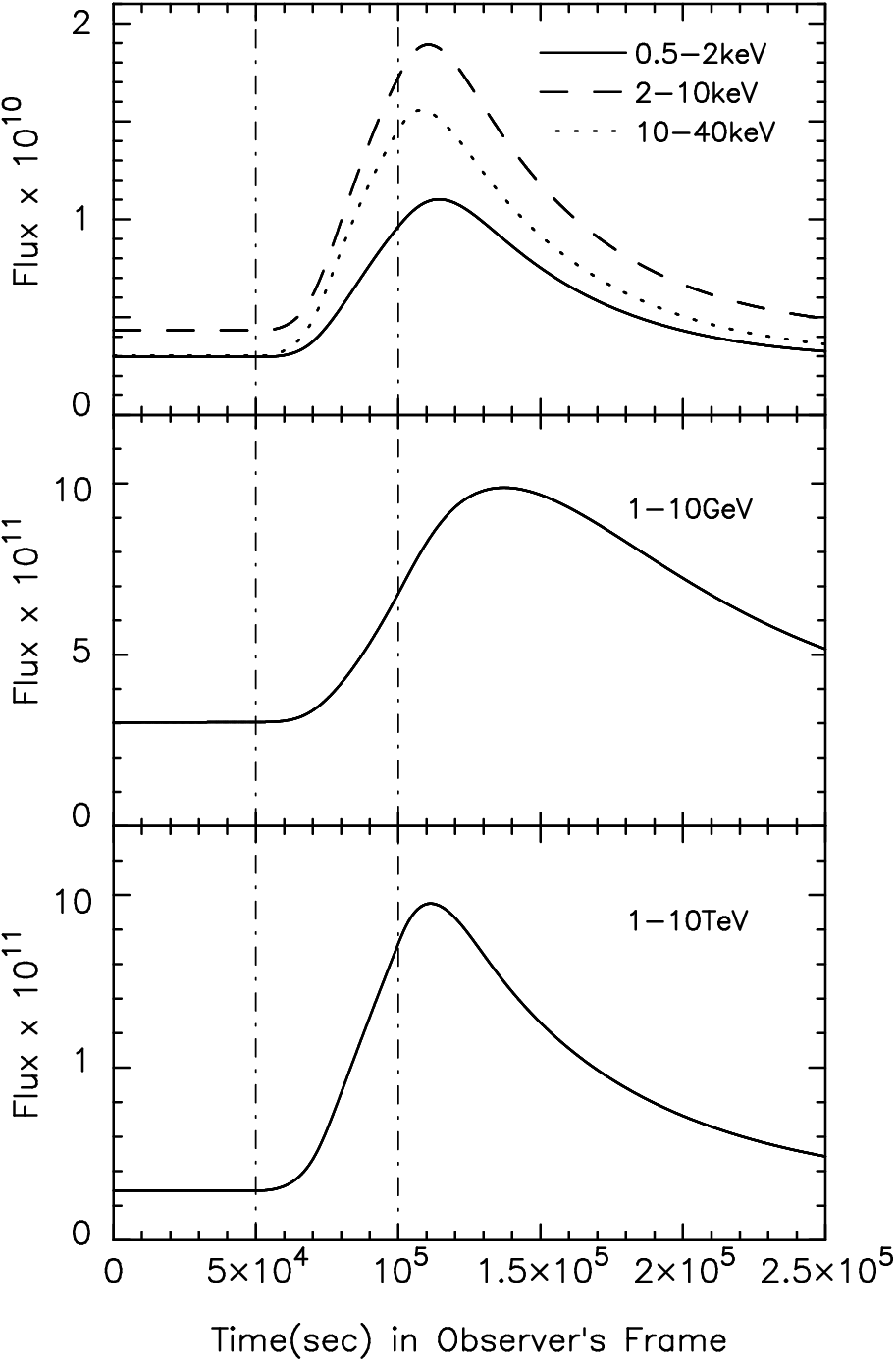}}

\figcaption{
Light curves for $t = 0 $ -- $5 R/c$ including a flare which occurs
during $t = R/c$ and $2 R/c$, indicated by the vertical dash-dotted lines. 
The fluxes are in units of ergs cm$^{-2}$ sec$^{-1}$.
\label{fig:light-flare}
}

% Figure 12
% \clearpage
%\epsfig{file=fig12.ps,height=15cm,width=15cm} 
\scalebox{0.8}[0.8]{\includegraphics{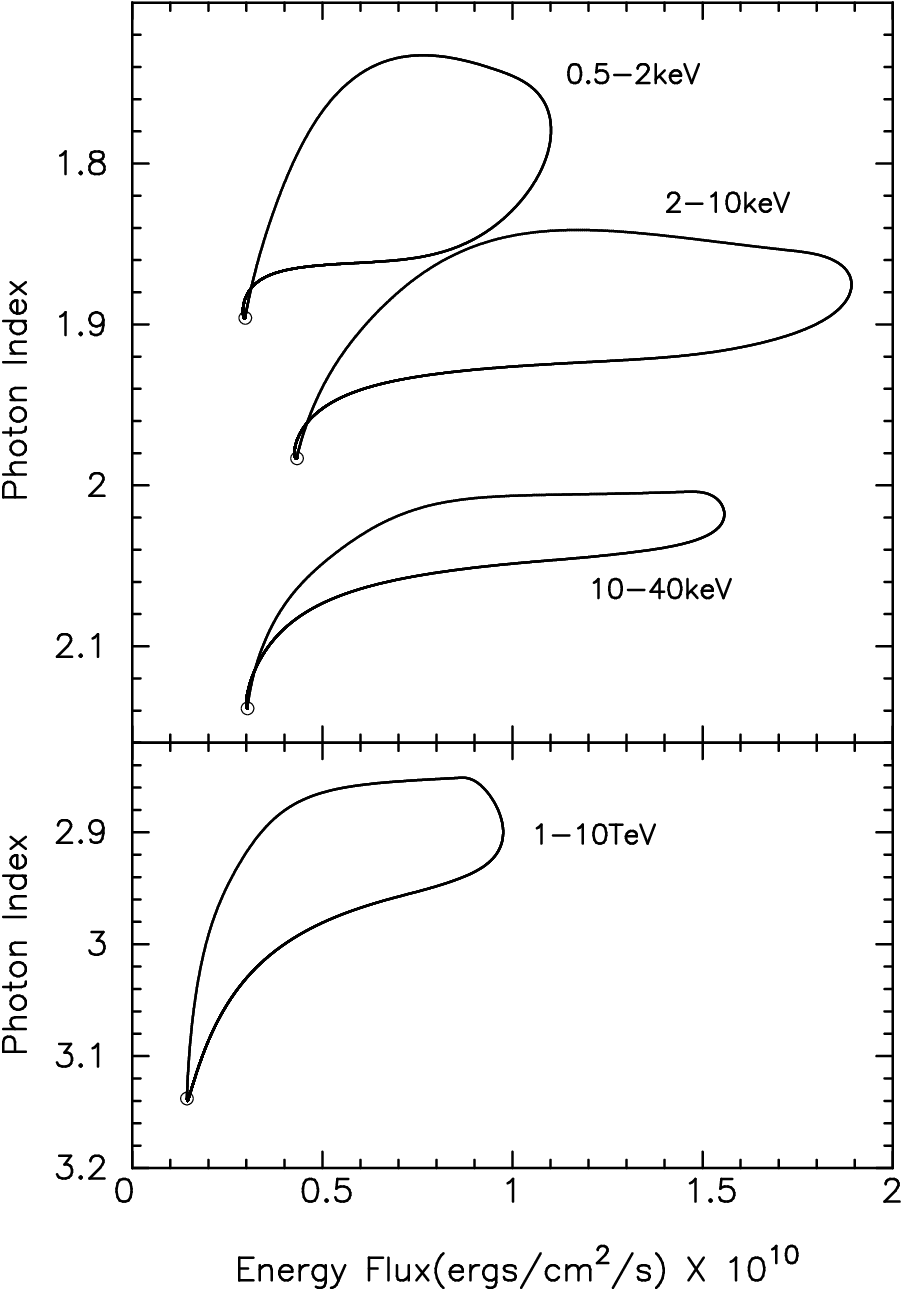}}

\figcaption{
Time evolution of the energy flux and the photon index associated with 
the flare shown in Figure \ref{fig:light-flare}; 
the trajectories start from a steady state (shown by open circles)
and rotate clockwise.  The evolution is calculated until $t = 10 R/c$.
\label{fig:alpha-flare}
}

\end{document}